\documentclass{article}
\usepackage{amssymb}
\usepackage{amsmath}
\def\R{{\mathbb R}}
\usepackage[dvipsnames]{xcolor}
\usepackage{tikz}
\usetikzlibrary{decorations.markings}
\tikzset{->-/.style={decoration={markings, mark=at position #1 with {\arrow{>}}},postaction={decorate}}}
\tikzset{-<-/.style={decoration={markings, mark=at position #1 with {\arrow{<}}},postaction={decorate}}}

\begin{document}

\title{Dynamics of the Selkov oscillator}

\author{Pia Brechmann and Alan D. Rendall\\
Institut f\"ur Mathematik\\
Johannes Gutenberg-Universit\"at\\
Staudingerweg 9\\
D-55099 Mainz\\
Germany}

\date{}

\maketitle

\begin{abstract}
A classical example of a mathematical model for oscillations in a biological
system is the Selkov oscillator, which is a simple description of glycolysis.
It is a system of two ordinary differential equations which, when expressed in
dimensionless variables, depends on two parameters. Surprisingly it appears 
that no complete rigorous analysis of the dynamics of this model has ever been 
given. In this paper several properties of the dynamics of solutions of the 
model are established. With a view to studying unbounded solutions a thorough 
analysis of the Poincar\'e compactification of the system is given. It is 
proved that for any values of the parameters there are solutions which tend to 
infinity at late times and are eventually monotone. It is shown that when the 
unique steady state is stable any bounded solution converges to the steady 
state at late times. When the steady state is unstable it is shown that for 
given values of the parameters either there is a unique periodic solution to 
which all bounded solutions other than the steady state converge at late times 
or there is no periodic solution and all solutions other than the steady state
are unbounded. In the latter case each unbounded solution which tends to 
infinity is eventually monotone and each unbounded solution which does not tend 
to infinity has the property that each variable takes on arbitrarily large and 
small values at arbitrarily late times.

\end{abstract}

\section{Introduction}

Glycolysis is a part of the process by which living organisms extract energy 
from sugar. It has been observed that under suitable circumstances the rate at 
which products of glycolysis accumulate shows oscillations in time although the 
input rate of sugar to the system is constant. Soon after these experimental
observations had been made Higgins \cite{higgins64} used mathematical modelling 
to obtain a deeper understanding of the process. Later it was observed by
Selkov \cite{selkov68} that the model of Higgins did not show sustained 
oscillations for biologically correct parameter values and he introduced an 
alternative mathematical model, which we refer to as the Selkov model. This 
system of two ordinary differential equations is the subject of what follows. 
(We comment further on the work of Higgins in the last section of the paper.)
It was shown in \cite{selkov68} that the Selkov model
exhibits a Hopf bifurcation and thus there exist parameter values for which it 
has a periodic solution. Results on the uniqueness and stability of periodic 
solutions of the Selkov model were obtained by d'Onofrio \cite{donofrio10}.

It turns out that solutions of the Selkov model can be unbounded and 
thus it becomes relevant to investigate the behaviour of the system near
infinity. This can be done by means of a Poincar\'e compactification.
This leads to two new steady states at infinity which are not hyperbolic.
It is relatively easy to determine the behaviour of solutions close to one of 
these steady states, which has a one-dimensional centre manifold, and this was 
done in \cite{selkov68}. The other steady state is more complicated since the 
linearization about that point vanishes identically and it was not treated
fully in \cite{selkov68}. To do so it is necessary to define some blow-ups of 
the singularity and this is done in what follows. It is suggested in 
\cite{selkov68}, on the basis of numerical calculations, that for certain 
parameter values there are unbounded solutions which approach infinity in an
oscillatory fashion. Analytically these might be interpreted as solutions 
which approach a heteroclinic cycle in (a suitable blow-up of) the Poincar\'e
compactification.  

Section \ref{basic} presents the basic facts about the model. In particular
it is shown that the model has a unique steady state for fixed 
values of the parameters and that there is a Hopf bifurcation. The first 
Lyapunov coefficient is calculated and shown to be negative. Thus the
bifurcation is generic and supercritical and there exist stable periodic 
solutions for parameter values close to those at the bifurcation. The subject 
of Section \ref{poincare} is the Poincar\'e compactification of the system.
In order to obtain a compactification where the dimension of the centre 
manifold of each steady state is at most one suitable blow-ups are carried
out. This allows the qualitative nature of the dynamics near the steady
states on the boundary of the compactification to be determined. In addition
to this local information it is shown that if there is a heteroclinic cycle
in the compactification it is asymptotically stable.

Section \ref{global} goes on to study the global properties of the phase 
portrait. It is shown that for any values of the parameters there are 
solutions which are unbounded at late times and eventually monotone. The
leading order asymptotics of these solutions is determined. The results of 
d'Onofrio on the uniqueness and stability of periodic solutions are reviewed 
and completed. It is proved that when the steady state is stable there exist 
no periodic solutions and all bounded solutions converge to the steady state
at late times. It is also proved that when the steady state is unstable
one of two mutually exclusive possibilities occurs. Either exactly one periodic 
solution exists and all bounded solutions converge to it at late times or no  
periodic solution exists and all solutions $(x(t),y(t))$ other than the steady 
state are unbounded. In the latter case for a given unbounded solution either 
$\lim_{t\to\infty}x(t)=\infty$ and $\lim_{t\to\infty}y(t)=0$ and $x(t)$ and $y(t)$
are eventually monotone or $\limsup_{t\to\infty}x(t)=\limsup_{t\to\infty}y(t)=
\infty$ and $\liminf_{t\to\infty}x(t)=\liminf_{t\to\infty}y(t)=0$. The final 
section discusses what remains to be discovered about the Selkov model and how 
this model relates to other models of glycolysis in the literature, in 
particular to a higher-dimensional model considered in \cite{selkov68}.

This paper is based in part on the MSc thesis of the first author 
\cite{brechmann17}.

\section{Basic facts}\label{basic}

The system considered in what follows is the system (II) of \cite{selkov68}. 
It is 
\begin{eqnarray}
&&\frac{dx}{d\tau}=1-xy^\gamma,\label{selkov1}\\
&&\frac{dy}{d\tau}=\alpha y(xy^{\gamma-1}-1).\label{selkov2}
\end{eqnarray}
It is written in dimensionless variables. The quantities $x$ and $y$ represent
dimensionless concentrations of ATP and ADP, respectively, while $\tau$ is a 
dimensionless time variable. The aim is to study the future evolution of 
solutions of these equations for which $x$ and $y$ are positive. The parameter 
$\alpha$ is a positive real number. The parameter $\gamma$ is a number greater 
than one and to avoid technical complications with differentiability 
properties it will be assumed here to be an integer.

The $x$-axis is an invariant manifold of the flow of the system 
(\ref{selkov1})-(\ref{selkov2}) and the vector field is directed towards
positive values of $x$ on the $y$-axis. For each fixed $\alpha$ there is a 
unique positive steady state and it satisfies $x=y=1$. Linearizing the system 
about this point leads to the Jacobi matrix
\begin{equation}
J=\left[
{\begin{array}{cc}
-1 & -\gamma\\
\alpha &\alpha (\gamma -1)\\
\end{array}}
\right].
\end{equation}
The determinant of $J$ is $\alpha$ which is always positive. Thus the stability 
of the steady state is determined by the trace of $J$, which is 
$\alpha(\gamma-1)-1$. Let $\alpha_0=\frac{1}{\gamma-1}$. Then for 
$\alpha<\alpha_0$ the trace of $J$ is negative and the steady state is
asymptotically stable while for $\alpha>\alpha_0$ the trace of $J$ is positive 
and the steady state is a source. The steady state is hyperbolic for all
$\alpha\ne\alpha_0$. For $\alpha=\alpha_0$ there is a pair of non-zero 
imaginary eigenvalues. If we consider the real part of the eigenvalues as
a function of $\alpha$ then it passes through zero when $\alpha=\alpha_0$ and 
its derivative with respect to $\alpha$ at that point is non-zero. Thus a 
Hopf bifurcation occurs. More information can be obtained by computing the
first Lyapunov number $\sigma$ of the bifurcation and it turns out that it 
is feasible to do this by hand using the formula given in \cite{perko01}.
The result is 
$\sigma=-\frac{3\pi(\gamma-1)^{\frac12}}{4}(\gamma^2(\gamma-1)+1)<0$.
Thus the Hopf bifurcation is non-degenerate and supercritical. This implies 
that for $\alpha$ slightly larger than $\alpha_0$ there exists a unique 
periodic solution close to the steady state and that this periodic solution 
is stable and hyperbolic. For $\alpha=\alpha_0$ the steady state is not 
hyperbolic but it is topologically equivalent to a hyperbolic sink, since
this is a property satisfied by all generic Hopf bifurcations. 

\section{The Poincar\'e compactification}\label{poincare}

The aim of this section is to investigate the ways in which the solutions of
the Selkov system can tend to infinity at late times. The first step in
doing this is to introduce two coordinate transformations, which we refer to
as Case 1 and Case 2, respectively. In Case 1 we define 
$Y=\frac{y}{x}$ and $Z=\frac{1}{x}$. These variables are 
appropriate for investigating the case where $x$ becomes large. The 
original variables can be recovered using the relations $x=\frac{1}{Z}$
and $y=\frac{Y}{Z}$. The result of the transformation is the system
\begin{eqnarray}
&&\frac{dY}{d\tau}=\frac{1}{Z^\gamma}(\alpha Y^\gamma+Y^{\gamma+1}
-\alpha YZ^\gamma-YZ^{\gamma+1}),\\
&&\frac{dZ}{d\tau}=\frac{1}{Z^{\gamma-1}}(Y^\gamma-Z^{\gamma+1}).
\end{eqnarray}
We introduce a new time variable $t$ by $\frac{dt}{d\tau}=Z^{-\gamma}$ and 
obtain the system
\begin{eqnarray}
&&\frac{dY}{dt}=\alpha Y^\gamma+Y^{\gamma+1}-\alpha YZ^\gamma-YZ^{\gamma+1},\\
&&\frac{dZ}{dt}=Y^\gamma Z-Z^{\gamma+2}.
\end{eqnarray}
In Case 2 we define $X=\frac{x}{y}$ and $Z=\frac{1}{y}$.
These variables are appropriate for investigating the case where $y$ 
becomes large.  The original variables can be recovered using the relations 
$x=\frac{X}{Z}$ and $y=\frac{1}{Z}$. The result of the transformation is the 
system
\begin{eqnarray}
&&\frac{dX}{d\tau}=\frac{1}{Z^\gamma}(Z^{\gamma+1}-X-\alpha X^2+\alpha XZ^\gamma)
,\\
&&\frac{dZ}{d\tau}=\frac{1}{Z^{\gamma-1}}(-\alpha X+\alpha Z^\gamma).
\end{eqnarray}
Introducing a new time variable as in Case 1 gives
\begin{eqnarray}
&&\frac{dX}{dt}=Z^{\gamma+1}-X-\alpha X^2+\alpha XZ^\gamma
,\\
&&\frac{dZ}{dt}=-\alpha XZ+\alpha Z^{\gamma+1}.
\end{eqnarray}
Case 2 is easier to analyse than Case 1 and so we begin with that. The system 
extends smoothly to the boundary where $Z=0$. This boundary is an invariant
manifold for the extended flow and there $X$ is decreasing except when $X=0$.
The origin in the new coordinates is a steady state, which we denote in what
follows by $P_1$. The linearization of the right hand side of the equations 
at $P_1$ has rank one and so there is a one-dimensional centre manifold there. 
The centre subspace is given by $X=0$ and the centre manifold is of the form 
$X=h(Z)$ where $h$ vanishes faster than linearly at the origin. The invariance
of the centre manifold under the flow implies that $\dot X=h'(Z)\dot Z$.
Putting this information into the evolution equations gives
\begin{equation*}
Z^{\gamma+1}-h(Z)-\alpha (h(Z))^2+\alpha (h(Z))Z^\gamma
=h'(Z)[-\alpha (h(Z))Z+\alpha Z^{\gamma+1}].
\end{equation*} 
If we expand this about $Z=0$ we see that each of the terms other than
$Z^{\gamma+1}$ and $h(Z)$ is negligible compared to one of these. It can be 
concluded that $h(Z)=Z^{\gamma+1}+o(Z^{\gamma+1})$. Substituting this 
relation into the evolution equation for $Z$ shows that on the centre manifold 
$\frac{dZ}{dt}=\alpha Z^{\gamma+1}+o(Z^{\gamma+1})$. Hence the flow on the centre 
manifold is away from the steady state. For $X=0$ the flow is into the 
positive region while between the line $Z=0$ and the centre manifold the flow 
is topologically equivalent to part of a hyperbolic saddle, as follows from
the reduction theorem \cite{kuznetsov95}. In particular, no solution 
starting in the positive region can tend to the point $P_1$ at late times.

We now turn to Case 1. The system extends smoothly to the boundaries where
$Y=0$ and $Z=0$ and these are invariant. The origin is a steady state.
Otherwise the variable $Z$ is decreasing when $Y=0$ and the variable $Y$ is 
increasing when $Z=0$. The linearization at the steady state is identically
zero and gives no information on the dynamics in a neighbourhood of that
point. To go further a blow-up procedure is applied to the steady state.
One way of doing this is to pass to polar coordinates. It turns out that a new
steady state is produced where the linearization is zero. Introducing 
polar coordinates a second time then leads to a successful analysis of 
the dynamics in the case $\gamma=2$, as was shown in \cite{brechmann17}.   
It seems, however, that for $\gamma>2$ this method becomes impractical. For 
this reason we instead use quasihomogeneous directional blow-ups 
\cite{dumortier06}. This allows the analysis to be carried out for general 
values of $\gamma$ in a unified way. It is necessary to do one such blow-up 
for each of the coordinates. They depend on two integers $\alpha$ and $\beta$ 
which are determined with the help of a Newton diagram \cite{dumortier06}. In 
the present case $\alpha=\gamma$ and $\beta=\gamma-1$.

The first blow-up uses the transformation $(Y,Z)\mapsto (\bar y,\bar z)$
with $(Y,Z)=(\bar y^\gamma,\bar y^{\gamma-1}\bar z)$. We introduce a new time 
coordinate by $\frac{ds}{dt}=\frac{1}{\gamma}\bar y^{\gamma^2-\gamma}$. Then the 
system becomes 
\begin{eqnarray}
&&\frac{d\bar y}{ds}=\alpha\bar y+\bar y^{\gamma+1}
-\alpha\bar y\bar z^\gamma-\bar y^{\gamma}\bar z^{\gamma+1},\\
&&\frac{d\bar z}{ds}=-\alpha(\gamma-1)\bar z
-(\gamma-1)\bar y^\gamma\bar z+\alpha(\gamma-1)\bar z^{\gamma+1}\nonumber\\
&&-\bar y^{\gamma-1}\bar z^{\gamma+2}+\gamma\bar y^{\gamma}\bar z.
\end{eqnarray} 
Evidently the origin of coordinates is a hyperbolic saddle which we denote
by $P_2$. There is one other steady state $P_3$ on the boundary where 
$(\bar y,\bar z)=(0,1)$. At $P_3$ there is a 
one-dimensional centre manifold, which will now be investigated. Introduce a 
new variable $w$ by the relation $\bar z=1+w$. We get the system
\begin{eqnarray}
&&\bar y'=-\alpha\gamma\bar y w-\bar y^\gamma (1+w)^{\gamma+1}
-\alpha\bar y[(1+w)^{\gamma}-1-\gamma w]
+\bar y^{\gamma+1},\nonumber\\
&&w'=\alpha\gamma(\gamma-1) w-\bar y^{\gamma-1}
-(\gamma-1)\bar y^{\gamma}(1+w)\\
&&+\alpha(\gamma-1)[(1+w)^{\gamma+1}-1-(\gamma+1)w]
-\bar y^{\gamma-1}[(1+w)^{\gamma+2}-1]\nonumber\\
&&+\gamma\bar y^{\gamma}(1+w).\nonumber
\end{eqnarray} 
Important properties of the centre manifold are collected in the following
lemma.

\noindent
{\bf Lemma 1} On the centre manifold of $P_3$ the relations 
$w=\nu_1\bar y^{\gamma-1}+o(\bar y^{\gamma-1})$ and 
$\bar y'=-\nu_2\bar y^{\gamma}+o(\bar y^{\gamma})$ hold
with $\nu_1=\frac{1}{\alpha\gamma(\gamma-1)}$ and 
$\nu_2=\frac{\gamma}{\gamma-1}$.

\noindent
{\bf Proof} Consider first the case $\gamma\ge 3$. To start with it will be
proved that 
\begin{equation}
w=O(\bar y^{\gamma-1}),\ \ \ w'=O(\bar y^{\gamma}).\label{wbound}
\end{equation}
For $\gamma\ge 3$ the centre subspace is given by $w=0$. Thus on the centre 
manifold $w=h(\bar y)=O(\bar y^2)$ with $h'(\bar y)=o(1)$ and hence
$\bar y'=O(\bar y^3)$. Differentiating the equation of the centre manifold 
gives $w'=h'(\bar y)y'$ and it follows that $w'=O(\bar y^3)$. This completes 
the proof of (\ref{wbound}) if $\gamma=3$. If $\gamma\ge 4$ we substitute 
the information obtained so far into the evolution equation for $w$ and 
obtain the statement that $w=O(\bar y^3)$. This implies that 
$\bar y'=O(\bar y^4)$ and that $w'=O(\bar y^4)$. Thus the proof of 
(\ref{wbound}) is complete if $\gamma=4$. For a general $\gamma$ we can repeat 
the argument just given until (\ref{wbound}) is obtained. Substituting the 
information obtained so far into the evolution equation for $w$ shows that 
\begin{equation}
w'=(\gamma-1)\left\{\alpha\gamma w-\frac{1}{\gamma-1}\bar y^{\gamma-1}
+O(\bar y^\gamma)\right\}.
\end{equation}
Combining this with (\ref{wbound}) gives the first assertion of Lemma 1.
Substituting the information already obtained back into the evolution equation 
for $\bar y$ gives 
\begin{equation}
\bar y'=-\alpha\gamma\bar y w-\bar y^\gamma 
+O(\bar y^{\gamma+1})
\end{equation}
which implies the second statement.

It remains to treat the case $\gamma=2$. In that case the centre subspace is 
of the form $\bar y=\mu w$ with $\mu=2\alpha$. It is 
convenient to introduce a new coordinate $v=\bar y-\mu w$, so that the centre 
subspace becomes $v=0$. The centre manifold is of the form 
$v=h(\bar y)=O(\bar y^2)$. Thus $w=\frac{1}{\mu}\bar y+O(\bar y^2)$. 
This already gives the first assertion of Lemma 1 in the case $\gamma=2$ 
since in that case $\nu_1=\frac{1}{\mu}$. Substituting the equation of the
centre manifold into the evolution equation for $\bar y$ gives
\begin{equation}
\bar y'=-2\alpha\bar y w-\bar y^2+O(\bar y^3)
\end{equation} 
which implies the second assertion in the case $\gamma=2$. $\blacksquare$

The second blow-up uses the transformation $(Y,Z)\mapsto (\bar y,\bar z)$
with $(Y,Z)=(\bar y\bar z^\gamma,\bar z^{\gamma-1})$. We introduce a new time 
coordinate by $\frac{ds}{dt}=\frac{1}{\gamma-1}\bar z^{\gamma^2-\gamma}$. Then 
the system becomes 
\begin{eqnarray}
&&\frac{d\bar y}{ds}=-\bar y^{\gamma+1}\bar z^{\gamma}
+\bar y\bar z^{\gamma-1}+\alpha(\gamma-1)\bar y^\gamma-\alpha(\gamma-1)\bar y,\\
&&\frac{d\bar z}{ds}=\bar y^\gamma\bar z^{\gamma+1}-\bar z^{\gamma}.
\end{eqnarray}
On the boundary $\bar z=0$ there is a steady state with coordinates $(1,0)$.
This is just another coordinate representation of the point $P_3$ and so it does
not need to be analysed further. We denote the steady state at the origin of 
these coordinates by $P_4$. At $P_4$ the $\bar z$-axis is a centre manifold.
Any other centre manifold at that point is of the form $\bar y=h(\bar z)$ where
the function $h$ vanishes to all orders for $\bar z\to 0$.

\begin{figure}[thb]
	\centering
	\begin{tikzpicture}[scale=1]
	\draw[thick,-<-=.27,->-=.78] (-1,0) arc(180:97:1cm and 1cm);
	\draw[thick,-<-=.5] (-4,4) arc(90:14.4:4cm and 4cm);
	\draw[thick,->-=.5] (-4,0)--(-1,0);
	\draw[thick] (-4,0)--(-4,4);
	\draw[ForestGreen,densely dotted,thick,->-=.6] (-1.15,0.95)--(-0.75,0.65);
	\draw[ForestGreen,densely dotted,thick,->] (-4,4) to [out=-90,in=130] (-3.8,3.5);
	\draw[thick,->] (-4.2,2.1)--(-3.8,1.9);
	\draw[ForestGreen,densely dotted,thick,->-=.5] (-1.4,0)--(-1,0);
	\draw[fill] (-0.13,0.99) circle(1pt) node[below right]{$P_2$};
	\draw[fill] (-0.75,0.65) circle(1pt) node[below right]{$P_3$};
	\draw[fill] (-1,0) circle(1pt) node[below right]{$P_4$};
	\draw[fill] (-4,4) circle(1pt) node[above left]{$P_1$};
	\draw (-0.9,3) node[]{$\gamma_1$};
	\draw (-1.3,0.5) node[]{$\gamma_2$};
	\end{tikzpicture}
	\caption{Poincar\'e compactification.}
	\label{FigPoincare}
\end{figure}
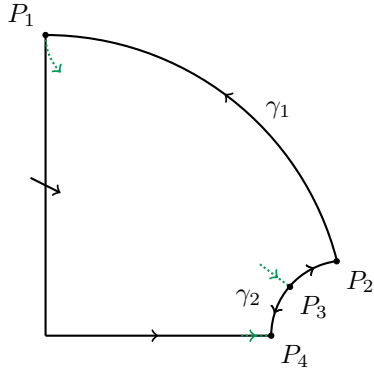

This analysis can be summarized as follows (cf. Fig.~\ref{FigPoincare}).

\noindent
{\bf Lemma 2} There is a smooth mapping $\phi$ of the closed positive quadrant 
into itself mapping the axes into themselves with the following properties.
The restriction of $\phi$ to the open quadrant is a diffeomorphism onto its
image. This image is a region whose closure is a compact set bounded by 
intervals $[0,x_0]$ and $[0,y_0]$ on the $x$- and 
$y$-axes and two smooth curves $\gamma_1$ and $\gamma_2$. The curve 
$\gamma_1$ joins the point $P_1=(0,y_0)$ with a point $P_2$ in the positive 
quadrant. The curve $\gamma_2$ joins the point $P_2$ with the point 
$P_4=(x_0,0)$. The image of the dynamical system can be rescaled so as to 
extend smoothly to the closure of the image of $\phi$ in such a way that 
$P_1$, $P_2$ and $P_4$ are steady states and $\gamma_1$, $\gamma_2$ and the 
image of the $x$-axis are invariant manifolds. There is precisely one 
further steady state $P_3$ on the boundary of the image of $\phi$ and it 
belongs to the interior of $\gamma_2$. 

The notation in Lemma 2 has been chosen to fit with that in the preceding 
analysis. Using that analysis we obtain the following picture of the flow of
the compactified system. Within the closure of the image of $\phi$ the points 
$P_1$, $P_2$ and $P_3$ are saddles while $P_4$ is a sink. The flow on 
$\gamma_1$ is towards 
$P_1$. The flow on $\gamma_2$ is away from $P_3$, i.e. towards $P_2$ and $P_4$. 
The flow on the horizontal axis is towards $P_4$. The centre manifold of $P_1$
enters the image of $\phi$ and the flow on it is away from $P_1$. The flow on 
the centre manifold of $P_3$ is towards $P_3$. It is conceivable that the 
centre manifold of $P_1$ might approach $P_3$, in which case it would coincide
with the centre manifold of $P_3$. In that case we say that there is an 
unbounded heteroclinic cycle. Note that a solution remains bounded towards the
future on any finite time interval. For it is obvious that $x$ remains bounded 
on such an interval and that $y$ remains bounded follows by considering the 
Poincar\'e compactification.

It will now be shown that if a heteroclinic cycle at infinity exists it is 
asymptotically stable. This is based on a study of the passage of a solution
close to the steady states belonging to the cycle. In fact it will be shown 
that this can be reduced to the study of the corresponding passages for 
some simplified dynamical systems, which will be examined first. The next 
lemma concerns the case of a hyperbolic steady state.

\noindent
{\bf Lemma 3} Consider the dynamical system $\dot x=-ax$, $\dot y=by$ for
positive constants $a$ and $b$. The solution which starts at the point $(1,y_0)$
reaches the line $y=1$ at the point $(x_1,1)$ with $x_1=(y_0)^{\frac{a}{b}}$.

\noindent
{\bf Proof} The dynamical system can be solved explicitly. Substituting the 
initial and final conditions into the solution gives an expression for the 
time $T$ required for the solution to go from the first to the second point. 
Substituting the expression for $T$ back into the solution gives the final 
result. $\blacksquare$

Treating a non-hyperbolic steady state requires the following more complicated
statement.

\noindent
{\bf Lemma 4} Consider the dynamical system 
\begin{eqnarray}
&&\dot x=-ax(1+\epsilon r_1(y,\epsilon)),\\
&&\dot y=by^k(1+\epsilon r_2(y,\epsilon)),
\end{eqnarray}
where $a>0$, $b>0$, $k\ge 2$ and $r_1$, $r_2$ are bounded. Suppose that the 
solution which starts at the point $(1,y_0)$ reaches the line $y=1$ at the 
point $(x_1,1)$. For any fixed $\eta>0$ and $\epsilon$ and $y_0$ sufficiently 
small the inequalities $x_1\le\exp (-[(1-\eta)a/(b(k-1))]y_0^{1-k})$ and 
$y_0\le C(-\log x_1)^{\frac{1}{1-k}}$ hold for a constant $C$.

\noindent
{\bf Proof} Suppose that the solution crosses the lines $x=1$ and $y=1$ for
$t=0$ and $t=T$, respectively. Given $\delta>0$ there exists $\epsilon_0>0$
such that $\epsilon\le\epsilon_0$ implies $|\epsilon r_1|\le\delta$ and 
$|\epsilon r_2|\le\delta$. It follows that 
$x_1\le e^{-a(1-\delta)T}$ and $y_0^{1-k}\le 1-b(1-k)(1+\delta)T$. Hence 
$T\ge \frac{1}{b(k-1)(1+\delta)}(y_0^{1-k}-1)$ and if $y_0$ is small enough
$x_1\le\exp\{-\frac{a(1-\delta)}{b(k-1)(1+\delta)^2}y_0^{1-k}\}$. For $\delta$
small enough $\frac{1-\delta}{(1+\delta)^2}\ge 1-\eta$ and we get the desired
inequality for $x_1$. On the other
hand $x_1\ge e^{-a(1+\delta)T}$ and $y_0^{1-k}\ge 1-b(1-k)(1-\delta)T$. Hence
$T\le \frac{1}{b(k-1)(1-\delta)}(y_0^{1-k}-1)$ and 
$x_1\ge\exp\{-\frac{a(1+\delta)}{b(k-1)(1-\delta)^2}y_0^{1-k}\}$. Hence
$y_0\le C(-\log x_1)^{\frac{1}{1-k}}$.

The reduction of the case of more general dynamical systems to these explicit
cases is a consequence of the following Lemma.

\noindent
{\bf Lemma 5} (i) Consider a two-dimensional dynamical system which has a 
steady state at a point $(x_*,y_*)$ which is a hyperbolic saddle, the 
eigenvalues of whose linearization at that point are $-a$ and $b$, where $a$
and $b$ are positive. Let $(x'_0,y'_0)$ and $(x'_1,y'_1)$ be points on the 
stable and unstable manifolds of $(x_*,y_*)$ which are sufficiently close to 
$(x_*,y_*)$. Fix curves $\beta_1$ and $\beta_2$ through $(x'_0,y'_0)$ and
$(x'_1,y'_1)$ which are transverse to the stable and unstable manifolds 
respectively. Let $(x_0,y_0)$ be a point on $\beta_1$ sufficiently close to,
but not equal to, $(x'_0,y'_0)$. By the Grobman-Hartman theorem this solution
intersects $\beta_2$ and let $(x_1,y_1)$ be the first such point it reaches.
Let $s_1$ and $s_2$ be any smooth coordinates on $\beta_1$ and $\beta_2$ which
vanish on the corresponding invariant manifolds and increase in the direction of
the points $(x_0,y_0)$ and $(x_1,y_1)$. Then for $s_1$ sufficiently small
the inequality $s_2\le C(s_1)^{\frac{a}{b}}$ holds for a positive constant $C$.

\noindent
(ii) Consider a two-dimensional dynamical system which has a steady state at a 
point $(x_*,y_*)$ whose centre and stable manifolds have dimension one. Suppose
that the non-zero eigenvalue is equal to $-a$, the flow on the centre manifold 
is away from $(x_*,y_*)$ and the leading term in the evolution 
equation along the centre manifold is $bs^k$ for a coordinate $s$ on that
manifold. Define points, curves and coordinates on the curves as in 
part (i) except that the unstable manifold is replaced by the centre 
manifold and the Grobman-Hartman theorem by the reduction theorem. Then for 
any $\eta>0$ the inequalities $s_2\le C\exp (-[(1-\eta)a/b(k-1)]s_1^{1-k})$
and $s_1\le C(-\log s_2)^{\frac{1}{1-k}}$ hold for $s_1$ sufficiently small and
a positive constant $C$.

\noindent
{\bf Proof} (i) It follows from Sternberg's theorem \cite{sternberg58}
that there is a smooth diffeomorphism which transforms the situation described 
to the model situation in Lemma 3 and this implies the desired statement. 
Note in particular that a suitable scaling of the coordinates ensures that
after the transformations the distance between the steady state and the points 
where the transverse sections cross the axis is one. Sternberg's 
theorem requires the hypothesis that there are no resonances. In the given 
case a resonance would mean an equation of the form $b=-na$ or $a=-nb$ for a 
positive integer $n$, which is clearly impossible. 

\noindent
(ii) In this case we must replace Sternberg's theorem by Takens' theorem
\cite{takens71}, allowing us to reduce the original system to the model system 
by a diffeomorphism of arbitrarily high finite differentiability. When there is 
only one non-zero eigenvalue the condition that there are no resonances is 
trivially fulfilled. In this case scaling the coordinate $y$ allows $\epsilon$
to be made arbitrarily small.

\noindent
{\bf Lemma 6} When for a given value of the parameter $\alpha$ in the Selkov
system a heteroclinic cycle at infinity exists this cycle is stable.

\noindent
{\bf Proof} Consider curves $\beta_i$ with the following properties. Let 
$\beta_1$ be a curve transverse to the unstable manifold of $P_2$ and 
sufficiently close to $P_2$. Similarly let $\beta_2$ be transverse to
the stable manifold of $P_1$ and close to $P_1$, $\beta_3$ transverse to
the centre manifold of $P_1$ and close to $P_1$, $\beta_4$ transverse to 
the centre manifold of $P_3$ and close to $P_3$, $\beta_5$ transverse to the 
unstable manifold of $P_3$ and close to $P_3$ and $\beta_6$ transverse to the
stable manifold of $P_2$ and close to $P_2$. For each $i$ let $s_i$ be a 
coordinate on $\beta_i$ which is zero on the relevant invariant manifold and
increases towards the interior of the image of $\phi$. Each solution which
crosses $\beta_1$ for a sufficiently small positive value of $s_1$ crosses each 
$\beta_i$ at some parameter value $s_i=f_{i-1}(s_{i-1})$. It then crosses 
$\beta_1$ again at some value $s_1=f_6(s_6)$. The aim is to obtain an estimate
for the composition $f$ of the $f_i$ which shows that the iterates of $f$ tend
to zero. The simplest mappings to estimate are $f_1$, $f_3$ and $f_5$. These 
maps are smooth at zero and so by Taylor's theorem there is a constant $C$
so that $f_1(s_1)\le C s_1$,  $f_3(s_3)\le C s_3$ and $f_5(s_5)\le C s_5$.
It is also relatively easy to estimate $f_6$ since $P_2$ is hyperbolic. Using
part (i) of Lemma 5 we get the estimate $s_1\le Cs_6^{\gamma-1}$. The mapping 
$f_2$ can be estimated using part (ii) of Lemma 5 with the result that 
$s_3\le Ce^{-cs_2^{-\gamma}}$ for a constant $c$. A similar estimate holds for
$s_4$. To estimate $f_4$ we use the second conclusion of Lemma 4. This 
gives $s_5\le C(-\log s_4)^{-\frac{1}{\gamma-1}}$. Composing 
$f_4\circ f_3\circ f_2$ and simplifying leads to an estimate of the 
form $s_5\le Cs_2^{\frac{\gamma}{\gamma-1}}$. Putting all these estimates together 
gives an inequality of the form $f(s_1)\le Cs_1^\gamma$. Hence if $s_1$ is
small the distance to the cycle is reduced by a fixed factor with each return 
and the cycle is stable.


\section{The global phase portrait}\label{global}

\begin{figure}
	\centering
	\begin{tikzpicture}[scale=0.6]
	\draw[thick] (-4,4) to [out=-60,in=165] (4,-2);
	\draw[thick] (-5,3) to [out=-45,in=170] (4,-1);
	\draw[fill] (-0.77,0.18) circle(2pt);
	\draw[thick,->] (2.7,-1)--(2.7,-0.4);
	\draw[thick,->] (-2.9,2.7)--(-3.5,2.7);
	\draw[thick,->] (-3.9,2.3)--(-3.9,1.7);
	\draw[thick,->] (1.8,-1.4)--(2.4,-1.4);
	\draw (4,-1.5) node[right]{\textcolor{gray}{$U_2$}};
	\draw (-4.5,3.5) node[above left]{\textcolor{gray}{$U_4$}};
	\draw (1,3) node[]{\textcolor{gray}{$U_3$}};
	\draw (-2,-1.5) node[]{\textcolor{gray}{$U_1$}};
	\draw (-4,4) node[above]{$N_{2,-}$};
	\draw (-5,3) node[left]{$N_{1,-}$};
	\draw (4,-2) node[below right]{$N_{2,+}$};
	\draw (4,-1) node[above right]{$N_{1,+}$};
	\draw[thick,->,gray] (-3.9,3.1)--(-4.2,2.8);
	\draw[thick,->,gray] (3,-1.45)--(3.25,-1.15);
	\draw[thick,->,gray] (0.5,1.5)--(0.1,1.75);
	\draw[thick,->,gray] (-1.5,-0.5)--(-1.1,-0.75);
	\end{tikzpicture}
	\caption{Nullclines.}
	\label{FigNullclines}
\end{figure}
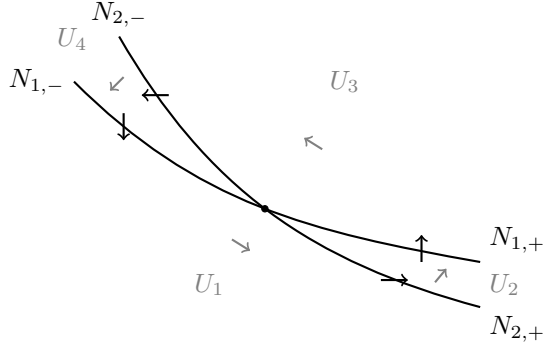

To understand the global phase portrait of the Selkov system it is helpful to
consider the nullclines $N_1$ and $N_2$ which are given by the equations
$xy^\gamma=1$ and $xy^{\gamma-1}=1$, respectively. The relevant geometry is 
illustrated in Figure~\ref{FigNullclines}.
The complement $U$ of 
$N_1\cap N_2$ has four connected components which can be distinguished by 
the signs of the right hand sides of the evolution equations. We denote them
by $U_1$, $U_2$, $U_3$ and $U_4$ for the combinations of signs $(+,-)$,
$(+,+)$, $(-,+)$ and $(-,-)$ respectively. The set $N_1\setminus N_2$ has 
two connected components $N_{1,+}$ and $N_{1,-}$ which are distinguished by the 
sign of $x-1$. Similarly $N_2\setminus N_1$ two connected components $N_{2,+}$ 
and $N_{2,-}$. A solution which starts at a point of $U_1$ for $t=t_0$ must
either stay in $U_1$ for $t\ge t_0$ or it must reach a point of $N_{2,+}$ after 
a finite time, after which it immediately enters $U_2$. If it stays in $U_1$ 
then either it is bounded and then it converges to $(1,1)$ for $t\to\infty$ 
or it is unbounded and in the latter case $\lim_{t\to\infty}x(t)=\infty$ and 
$\lim_{t\to\infty}y(t)=0$. A solution which starts in $U_2$ for $t=t_0$ must
reach a point of $N_{1,+}$ after a finite time, after which it immediately 
enters $U_3$. If a solution which starts in $U_3$ for $t=t_0$ were unbounded
while remaining in $U_3$ then it would approach $P_1$ in the Poincar\'e
compactification for $t\to\infty$ and this has already been ruled out in the
last section. Thus it must either converge to $(1,1)$ as $t\to\infty$ or 
reach $N_{2,-}$ after a finite time, after which it immediately enters $U_4$.
A solution which starts in $U_4$ for $t=t_0$ must reach a point of $N_{1,-}$ 
after a finite time, after which it immediately enters $U_1$. Putting all 
this information together we get the following result

\noindent
{\bf Lemma 7} Each solution of the Selkov system has one of the following
behaviours in the future. 

\noindent
(i) it converges to $(1,1)$ for $t\to\infty$ while staying in $U_1$ or $U_3$

\noindent
(ii) $\lim_{t\to\infty}x(t)=\infty$ and $\lim_{t\to\infty}y(t)=0$

\noindent
(iii) it cycles indefinitely between the regions $U_i$.

A similar analysis can be done in the past time direction. Note that it is not
possible that a solution is always in $U_3$ before a certain time because no
solution can approach the curve $\gamma_2$ in the Poincar\'e compactification
towards the past. Thus we have the following result

\noindent
{\bf Lemma 8} Each solution of the Selkov system has one of the following
behaviours in the past.

\noindent
(i) it converges to $(1,1)$ for $t\to -\infty$ while staying in $U_1$ or $U_3$

\noindent
(ii) it starts on the $y$-axis

\noindent
(iii) $\lim_{t\to -\infty}x(t)=0$ and $\lim_{t\to -\infty}y(t)=\infty$

\noindent
(iv) it cycles indefinitely between the regions $U_i$. 

If a solution cycles indefinitely in the future then it passes through points 
of the form $(1,y)$ with $y<1$ at an infinite increasing sequence of times 
$t_i$. Let $y_i=y(t_i)$. It follows from Poincar\'e-Bendixson theory that the 
sequence $y_i$ is monotone. If the sequence is constant then the solution is 
periodic. Otherwise the sequence is strictly monotone. If it is monotone 
decreasing then it must tend to some $y_{-,*}>0$. That this number is strictly 
positive follows
from the fact that in the Poincar\'e compactification all solutions which start
with $x=1$ and $y$ sufficiently small converge to $P_4$ for $t\to\infty$. 
Poincar\'e-Bendixson theory further shows that either $(1,y_{-,*})$ lies on a 
periodic solution or on a heteroclinic cycle at infinity. If the sequence
$y_n$ is monotone increasing then $y_{-,*}$ lies on a periodic solution.
This discussion tells us that to classify the asymptotic behaviour of all
solutions of the Selkov system for given values of $\gamma$ and $\alpha$ it 
suffices to know how many periodic solutions there are, whether there is a 
heteroclinic cycle at infinity and what the stability properties of the 
periodic solutions and the heteroclinic cycle are. Note that every periodic 
solution must contain a point of the form $(1,y)$ with $y<1$.

For any values of the parameters in the Selkov system there are unbounded 
solutions where $x$ gets large and $y$ gets small at late times and which
have an asymptotic behaviour which can be determined. This is the content
of the following theorem. These are exactly the solutions which belong to
Case (ii) in Lemma 7.

\noindent
{\bf Theorem 1} There exists a positive number $\epsilon>0$ such that any 
solution of the Selkov system with initial data $x(0)=x_0$ and $y(0)=y_0$ which 
satisfies $x_0>\epsilon^{-1}$ and $x_0y_0^\gamma<\epsilon$ has the late-time 
asymptotics
\begin{eqnarray}
&&x(\tau)=\tau (1+o(1)),\\
&&y(\tau)=y_1 e^{-\alpha\tau}(1+o(1)).
\end{eqnarray}
for a constant $y_1$. There exists a solution, unique up to time translation, 
which has the asymptotic behaviour
\begin{eqnarray}
&&x(\tau)=\tau (1+o(1)),\\
&&y(\tau)=\tau^{-\frac{1}{\gamma-1}}(1+o(1)).
\end{eqnarray}

\noindent
{\bf Proof} Any solution which starts close to the point $P_4$ converges to
that point as $t\to\infty$. Using this information in the evolution equations 
for $\bar y$ and $\bar z$ allows them to be integrated in leading order. First 
we have 
\begin{equation}
\frac{d\bar z}{ds}=-\bar z^\gamma(1+o(1))\label{asymp1}
\end{equation}
and hence
\begin{equation}
\bar z(s)=[(\gamma-1)s]^{-\frac{1}{\gamma-1}}(1+o(1)).\label{zleading}
\end{equation}
Putting this information back into the evolution equation allows the error term
$o(1)$ to be improved to $O(s^{-\frac{1}{\gamma-1}})$. The relation 
\begin{equation}
\frac{d\bar y}{ds}=\bar y(-(\gamma-1)\alpha+o(1))\label{asymp2}
\end{equation}
shows that $\bar y$ decays exponentially as $t\to\infty$. Using this fact
and substituting the improved version of (\ref{zleading}) into the evolution 
equation for $\bar y$ gives
\begin{equation}
\frac{d\bar y}{ds}=\bar y\left[-(\gamma-1)\alpha+\frac{1}{\gamma-1}s^{-1}
+q(s)\right]
\label{asymp3}
\end{equation}
where the function $q$ is integrable on intervals of the form $[s_0,\infty)$.
Integrating this last relation gives 
\begin{equation}
\bar y(s)=As^{\frac{1}{\gamma-1}}e^{-\alpha(\gamma-1)s}(1+o(1))\label{yleading}
\end{equation}
for a constant $A$. It follows from the estimates obtained up to now that 
the image of each such solution is tangent to the $\bar z$-axis at $P_4$ and is 
therefore a centre manifold at $P_4$. 
It remains to transform these formulae from the variables 
$(\bar y,\bar z,s)$ to the variables $(x,y,\tau)$. Note that 
$\frac{ds}{dt}=\frac{1}{\gamma-1}Z^\gamma$ and $\frac{d\tau}{dt}=Z^\gamma$.
The coordinate $s$ is only defined up to an additive constant and if this
constant is chosen appropriately it follows that $\tau=(\gamma-1)s$. Using 
the expressions (\ref{asymp1}) and (\ref{asymp2}) and the definitions of 
the coordinate transformations gives the first part of the theorem. The 
solution mentioned in the second part of the theorem is a solution on the 
centre manifold of $P_3$. Integrating the equation for $\bar y$ in Lemma 1
gives $\bar y(s)=(\nu_2(\gamma-1)s)^{-\frac{1}{\gamma-1}}(1+o(1))
=(\gamma s)^{-\frac{1}{\gamma-1}}(1+o(1))$. Putting this
into the equation for $w$ given there leads to
$w(s)=\nu_1(\gamma s)^{-1}(1+o(1))$. Now 
$\frac{ds}{dt}=\gamma^{-1}(\gamma s)^{-\gamma}(1+o(1))$. On the other hand
$\frac{d\tau}{dt}=Z^{\gamma}=(\gamma s)^{-\gamma}(1+o(1))$. Thus we can choose
$s$ so that $\tau=\gamma s$. Using the definitions of the coordinate
transformation gives the second part of the theorem. $\blacksquare$ 

To obtain more information about periodic solutions it is helpful to do the
following changes of variables. Introduce a time variable by 
$\frac{dt}{d\tau}=y^\gamma$ and define $v=\frac{dy}{dt}$. Let 
$f(y)=1-\alpha(\gamma-1)y^{-\gamma}$ and $g(y)=\alpha (y-1)y^{-\gamma}$.
Denote by $F$ the primitive of $f$ which vanishes for $y=1$. Introduce 
variables by $\tilde x=y-1$ and $\tilde y=-v-F(y)$. Dropping the tildes gives
the system
\begin{eqnarray}
&&\dot x=-y-F(x),\label{lienard1}\\
&&\dot y=g(x).\label{lienard2}
\end{eqnarray}
We are interested in solutions of this for which $x>-1$ and where there is no 
restriction on $y$. It will be shown that under suitable circumstances the 
system (\ref{lienard1})-(\ref{lienard2}) defined on a region of the form 
$(x_0,\infty)\times\R$ has at most one periodic solution and that if such a 
solution exists it is asymptotically stable. This implies a corresponding 
result for the Selkov system. Next we discuss some auxiliary results used in 
the proof of these statements.

\noindent
{\bf Theorem 2} Suppose that the system (\ref{lienard1})-(\ref{lienard2}) 
satisfies the following conditions:

\noindent
(i) $xg(x)>0$ when $x\ne 0$ and $g'(0)>0$

\noindent
(ii) $F(0)=0$ and $f(0)<0$ where $f(x)=F'(x)$

\noindent
(iii) There exists a real number $\alpha$ such that the function
$f_1(x)=f(x)+\alpha g(x)$ has zeroes $x_1<0$ and $x_2>0$ with
$f_1(x)\le 0$ for $x_1<x<x_2$ and there exists 
$x_*$ with $f_1(x_*)>0$ and $(x_*,0)$ in the interior of $L_1$. 

\noindent
(iv) All the limit cycles are contained in the interval $x_3\le x\le x_4$, where
$x_3<x_1<0<x_2<x_4$ and the function $f_1/g$ does not decrease for 
$x_3\le x\le x_1$ and $x_2\le x\le x_4$.

\noindent
(v) all the limit cycles contain the interval $x_1\le x\le x_2$ on the $x$-axis.

\noindent
Then the system has at most one limit cycle and if one exists it is stable.

This theorem is closely related to a result stated in Kuang and Freedman 
\cite{kuang88}, which is in turn related to a result of Cherkas and Zhilevich 
\cite{cherkas70}. The differences between Theorem 2 and Theorem 3.1 in
\cite{kuang88}, apart from the different notation, are as follows. For 
simplicity the function $\phi(y)$ in \cite{kuang88} has been set to $y$ and 
the constant $\beta$ has been set to zero. The condition (4) of the theorem of 
\cite{kuang88} has been replaced by the condition (4)${}'$. The condition 
$g'(0)>0$ has been added since without it the proof in \cite{kuang88} is
incomplete. The conditions that $x_1$ and $x_2$ are simple zeroes have been 
replaced by the assumption concerning the existence of the point $x_*$.
This is because we could not see how to prove that Theorem 3.2 of \cite{kuang88}
follows from Theorem 3.1, a claim made without proof in the paper.
  
\noindent
{\bf Proof of Theorem 2} We give only an outline of the proof. More details 
can be found in \cite{kuang88} and \cite{brechmann17}. The unique steady state 
of the system is at the origin. It follows from the condition $f(0)<0$ that 
this steady state is unstable. The proof of the theorem is by contradiction 
and we assume that there exist two limit cycles $L_1$ and $L_2$. It is 
necessarily the case that the steady state is in the interior of the Jordan 
curves defined 
by the solutions and that one of them is in the interior of the other. Suppose 
that $L_1$ is in the interior of $L_2$. To examine the stability of the 
solutions the Poincar\'e stability criterion \cite{perko01} will be used. It 
involves the quantity obtained by integrating the divergence of the vector 
field defining the dynamical system over the solution. We call this the 
Poincar\'e quantity. Let $A$ and $A_1$ be the points on $L_1$ and $L_2$ with 
$x=x_2$ and $y>0$ and let $D$ and $D_1$ be the points with with $x=x_2$ and 
$y<0$. Let $E$ and $E_1$ be the points on $L_2$ with the same $y$ coordinates 
as $A$ and $D$. Integrate the differential form $(f_1/g)dy$ over the boundary 
of the region $R_1$ bounded by the line segments $AE$ and $DE_1$ and the parts 
of $L_1$ and $L_2$ joining $A$ with $D$ and $E$ with $E_1$. This results in 
the sum of two integrals along the curved parts of the boundary. On the other 
hand, by Stokes' theorem this integral is equal to the integral of a 
non-negative quantity over $R_1$. Thus we get an inequality relating the 
integrals of $f_1$ along certain parts of $L_1$ and $L_2$. We can do a similar 
construction starting from $x=x_1$ with points $B,B_1,C,C_1,K,K_1$ 
corresponding to $A,A_1,D,D_1,E,E_1$. Next we replace $(f_1/g)dy$ in this 
construction by $f_1/(y+F(x))dx$ and integrate over regions with corners 
$C,D,C_1,D_1$ and $A,B,A_1,B_1$. Putting these computations together shows 
that the integral of $f_1$ over $L_2$ is equal to its integral over $L_1$ plus 
contributions from the parts of $L_2$ joining $E$ with $A_1$, $B_1$ with $K$, 
$K_1$ with $C_1$ and $D_1$ with $E_1$. In fact these extra contributions are 
non-negative. This is because $f_1$ is non-negative on the relevant intervals. 
Moreover it is positive on the interval $[x_*,x_4]$, due to the monotonicity 
of $f_1/g$. It follows that the integral of $f_1$ over $L_2$ is strictly 
greater than its integral over $L_1$. It can be shown that the integrals of $g$
over $L_1$ and $L_2$ are zero so that the integral of $f$ over $L_2$ is 
strictly greater than its integral over $L_1$. The divergence of the 
vector field defining the system is equal to $-f$ and so the Poincar\'e 
quantity for $L_2$ is less than that for $L_1$. Since $g'(0)>0$ the
steady state is a hyperbolic source. Hence there must be an innermost periodic 
solution and there must be a solution starting near the steady state and 
converging to this periodic solution for $t\to\infty$. As a consequence it is 
a limit cycle and we choose it as $L_1$. It follows that $L_1$ is stable from 
the inside and that the Poincar\'e quantity is non-positive and hence the 
integral of $-f$ over $L_1$ also non-positive. Hence the Poincar\'e quantity 
for $L_2$ must be negative. If $L_1$ were stable on the outside we could choose 
$L_2$ to be the closest limit cycle outside it. $L_2$ would have 
to be unstable on the inside, a contradiction. Thus $L_1$ is semistable 
(stable on one side and unstable on the other). It can be shown by perturbing 
the system using the method of rotated vector fields \cite{duff53} that if 
this were the case there would be a system where the above contradiction is 
obtained. It follows that no second limit cycle can exist and if one limit 
cycle does exist it is stable. $\blacksquare$

This theorem can be used to prove another where the opaque condition 
involving the function $f_1$ is removed from the hypotheses. It is a 
modification of a theorem of Zhang \cite{zhang86}.

\noindent
{\bf Theorem 3} Suppose that the system (\ref{lienard1})-(\ref{lienard2}) 
satisfies the conditions (i) and (ii) of Theorem 2 and that all
limit cycles are contained in the interval $a<x<b$ where $a<0<b$ and 
$f(x)/g(x)$ is non-decreasing when $x$ increases in $a<x<0$ and $0<x<b$.
Then the conclusion of Theorem 2 holds.

\noindent
{\bf Proof} The condition that $f_1/g$ is non-decreasing is equivalent to the
condition that $f/g$ is non-decreasing. The requirement on the region where
this function is non-decreasing in the hypotheses of Theorem 3 implies that 
in the hypotheses of Theorem 2. We need to show that $\alpha$ can be chosen
so that $f_1$ has zeroes with the properties required in condition (iii) of 
Theorem 2. Following \cite{zhang86} let $\tilde x$ be the smallest value of 
$x$ attained at a point of $L_1$ and define 
$\alpha=-\frac{f(\tilde x)}{g(\tilde x)}$. Then $f_1(\tilde x)$ is zero by 
construction and $\tilde x$ is a zero of $f_1$ with $\tilde x<0$. Let $x_1$ be 
the largest value of $x$ for which $f_1(x)=0$ for $\tilde x\le x\le x_1$. 
Since $f_1(0)<0$ we can conclude that $x_1<0$. We will now show that $f_1$ has 
a second zero $x_2>x_1$ which is in the interior of $L_1$. If this were not 
the case then $L_1$ would lie completely in the region where $f_1$ is 
non-positive, as will now be shown. For $\frac{f_1(x_1)}{g(x_1)}=0$ and we 
have assumed that 
$\frac{d}{dx}\left(\frac{f_1(x)}{g(x)}\right)$ is non-negative on the 
interval $[x_1,0)$. Thus $\frac{f_1(x)}{g(x)}\ge 0$ and $f_1(x)\le 0$ on 
that interval. At the origin $\frac{f_1(x)}{g(x)}<0$ and so in fact 
$\frac{f_1(x)}{g(x)}$ is negative on an interval beginning at $x_1$ and 
extending beyond $x=0$. This establishes the claim concerning $L_1$. It
can be concluded that the integral of $f_1$ over $L_1$ is negative. For
under the given assumptions $f_1$ is non-positive everywhere on $L_1$ and 
negative at some point of $L_1$. The Poincar\'e criterion then implies that 
$L_1$ is unstable in the interior. This contradicts the fact that there are 
solutions starting near the steady state which converge to $L_1$. Thus in 
reality $f_1$ has a zero $x_2>0$ in the interior of $L_1$. In fact $f_1$ must 
become positive for some $x_*>0$ with $(x_*,0)$ in the interior of $L_1$ since 
otherwise the contradiction would still occur. Choosing $x_2$ to be the smallest
$x>0$ for which $f_1$ is negative on $(0,x_2)$ ensures that the hypotheses of
Theorem 3 are satisfied. $\blacksquare$





Theorem 3 can be used to obtain a result about the Selkov system following 
d'Onofrio.

\noindent
{\bf Theorem 4} \cite{donofrio10} If the steady state in the Selkov model is
unstable for a given value of $\alpha$ then there is at most one limit cycle
and if one exists it is asymptotically stable.

In his proof of this theorem d'Onofrio cites the Theorem of Kuang and Freedman.
However the theorem he formulates (Proposition 1 of his paper) is not equivalent
to the corresponding result stated by Kuang and Freedman (Theorem 3.2 of their
paper). In \cite{kuang88} the assumptions include the condition that 
$\phi(y)+F(x)$ is defined for all $x\in (-\infty,\infty)$. In the case 
$\phi(y)=y$ of interest here this means that $F$ would have to be defined on
the whole real line. This condition is not assumed in d'Onofrio's Proposition
1 and indeed it does not hold in the situation of the application to the 
Selkov model. It is this apparent gap which motivates our discussion of 
Theorem 2 and Theorem 3 above. That discussion shows that the extra assumption
is not necessary. When this has been clarified a calculation done by d'Onofrio
showing the monotonicity of $f/g$ in the case of the Selkov model suffices
to show that Theorem 4 follows from Theorem 3.

Theorem 4 can be used to obtain information about the long-time behaviour of
solutions in the case that the steady state is unstable. Using 
Poincar\'e-Bendixson theory it follows that the $\omega$-limit set in the 
Poincar\'e compactification of any solution which cycles indefinitely in the 
future (case (iii) of Lemma 7) is either the unique periodic solution
or the heteroclinic cycle at infinity. We know as a result of Lemma 6 and 
Theorem 4 that both the periodic solution and the heteroclinic cycle are 
stable whenever they exist. If both existed then by continuity there 
would have to be a periodic solution between them and this would contradict
Theorem 4. Thus we obtain the following result

\noindent
{\bf Theorem 5} In the case $\alpha>\alpha_0$ exactly one of the following 
three situations occurs.

\noindent
(i) The centre manifolds of $P_1$ and $P_3$ coincide so that there is a 
heteroclinic cycle at infinity. Any solution which starts below this centre
manifold converges to $P_4$ as $t\to\infty$ while any solution other than the 
steady state which starts above this manifold converges to the heteroclinic 
cycle at infinity as $t\to\infty$.

\noindent
(ii) The centre manifolds of $P_1$ and $P_3$ do not coincide. There exists a 
unique periodic solution. Any solution which starts below the centre manifold of
$P_3$ converges to $P_4$ as $t\to\infty$ while any solution other than the 
steady state which starts above this manifold converges to the periodic 
solution as $t\to\infty$.  

\noindent
(iii) The centre manifolds of $P_1$ and $P_3$ do not coincide. Any solution
other than the steady state which does not lie on the centre manifold of 
$P_3$ converges to $P_4$ as $t\to\infty$. 

We next turn to the case $\alpha\le\alpha_0$. For this case d'Onofrio 
proved a result (Proposition 5.3 of his paper) as an application of a Theorem 
of Hwang and Tsai \cite{hwang05}. It was shown in \cite{brechmann17} that one 
of the conditions in his result is satisfied automatically for 
$\alpha\le\alpha_0$, thus showing that the result can be generalized. We do
so in Theorem 7 below. The following theorem is closely related to Theorem 2.1 
of \cite{hwang05}.

\noindent
{\bf Theorem 6} Suppose that in the system (\ref{lienard1})-(\ref{lienard2})
defined on the region $(r_1,r_2)\times\R$ the following conditions hold:

\noindent
(i) there exists $\lambda\in (r_1,r_2)$ such that $g'(\lambda)>0$ and 
$(x-\lambda)g(x)>0$ for all $x\in (r_1,r_2)\setminus\{\lambda\}$.

\noindent
(ii) there exist constants $a,b\in\R$ such that the function
$f(x)+ag(x)+bg(x)F(x)$ is non-negative on $(r_1,r_2)$ and not identically
zero.

\noindent
Then the system has no periodic solutions.

The relation of this statement to that of Theorem 2.1 of \cite{hwang05} is
that $y$ has been replaced by $-y$, the function $\phi$ has been taken to 
be equal to one and the function $\pi(y)$ equal to $y$. Under these 
circumstances conditions (A1), (A2) and (A4) of the theorem of 
\cite{hwang05} are satisfied automatically and the remaining conditions
reduce to (i) and (ii) above. In the case of the Selkov system the interval
can be chosen to be $(-1,\infty)$ and condition (i) is satisfied by 
$\lambda=0$. 
Thus it only remains to consider the condition (ii) and in fact it suffices to 
choose $b=0$.

Substituting the values of $f$ and $g$ in the case of the Selkov system into
the positivity condition in (ii) gives
\begin{equation}
(x+1)^\gamma+a\alpha (x+1)+\alpha(1-\gamma-a)\ge 0.
\end{equation}
Let $\beta=\alpha (\gamma-1)$, so that the stability condition is $\beta\le 1$.
Choose $a=-\gamma (\gamma-1)$. Then 
\begin{eqnarray}
&&(x+1)^\gamma+a\alpha (x+1)+\alpha(1-\gamma-a)\nonumber\\
&&=(x+1)^\gamma-\alpha\gamma (\gamma-1)(x+1)+\alpha (\gamma-1)^2\\
&&=(x+1)^\gamma-\beta\gamma (x+1)+\beta(\gamma-1).
\end{eqnarray}
The last function has a global minimum in $(0,\infty)$. This occurs at the 
point $x+1=\beta^{\frac{1}{\gamma-1}}$. Substituting this back into the function
shows that the minimum is $(\gamma-1)\beta (1-\beta^{\frac{1}{\gamma-1}})\ge 0$. 
Thus this choice of $a$ has the desired property and Theorem 6 can be 
applied to the Selkov system.

\noindent
{\bf Theorem 7} If the positive steady state of the Selkov sytem is stable 
then there are no periodic orbits and every solution either tends to the 
positive steady state or tends to $P_3$ or $P_4$ at late times.

\noindent
{\bf Proof} The first statement follows from Theorem 6. There cannot be a 
heteroclinic cycle at infinity since the fact that both the cycle and the
steady state are stable would mean that there would have to be a periodic
solution between them. The only remaining possibilities for the $\omega$-limit
set of a solution are those listed in the theorem. 

\section{Further remarks}

The main question which has been left open by the analysis in this paper is
whether there are parameter values for which there are solutions of the 
Selkov model which exhibit unbounded oscillations. This is the question, whether
case (i) in Theorem 5 ever occurs. To prove this it would suffice to show that 
there is a value of $\alpha$ for which cases (ii) and (iii) of Theorem 5 are 
ruled out. To rule out case (ii) for a given value of $\alpha$ it would be 
enough to show that the system admits a Lyapunov function or a Dulac function. 

The model (\ref{selkov1})-(\ref{selkov2}) which we have studied in this paper 
was originally derived by Selkov from a five-dimensional model, the system (4)
in \cite{selkov68}, by a singular limiting process. This limit will now be 
considered. The variables $s_1$, $s_2$, $x_1$, $x_2$ and $e$ in the larger
model are the concentrations of the substrate ATP, the product ADP,
the activated enzyme, the complex between the activated enzyme and the substrate
and the inactive enzyme, respectively. With some assumptions about the nature 
of these reactions and assuming mass action kinetics the following system of 
equations is obtained.
\begin{eqnarray}
&&\frac{ds_1}{dt}=v_1-k_1s_1x_1+k_{-1}x_2,\\
&&\frac{ds_2}{dt}=k_2x_2-\gamma k_3s_2^\gamma e+\gamma k_{-3}x_1-ks_2,\\
&&\frac{dx_1}{dt}=-k_1s_1x_1+(k_{-1}+k_2)x_2+k_3s_2^\gamma e-k_{-3}x_1,\\
&&\frac{dx_2}{dt}=k_1s_1x_1-(k_{-1}+k_2)x_2,\\
&&\frac{de}{dt}=-k_3s_2^\gamma e+k_{-3}x_1.
\end{eqnarray}
Note that $e_0=e+x_1+x_2$ is a conserved quantity (total amount of enzyme) and 
this can be used to eliminate $e$ from the first four evolution equations and 
discard the evolution equation for $e$. This reduces the system to four 
equations.

Next dimensionless variables are introduced by defining 
$\sigma_1=\frac{k_1s_1}{k_{-1}+k_2}s_1$, 
$\sigma_2=\left(\frac{k_3}{k_{-3}}\right)^{\frac{1}{\gamma}}s_2$, 
$u_1=\frac{x_1}{e_0}$, $u_2=\frac{x_2}{e_0}$, 
$\theta=\frac{e_0k_1k_2}{k_{-1}+k_2}t$. This leads to the system
\begin{eqnarray}
&&\frac{d\sigma_1}{d\theta}=\nu-\frac{k_2+k_{-1}}{k_2}u_1\sigma_1
+\frac{k_{-1}}{k_2}u_2,\label{selkove1}\\
&&\frac{d\sigma_2}{d\theta}=\eta\left(u_2-\gamma\frac{k_{-3}}{k_2}
\sigma_2^\gamma(1-u_1-u_2)+\gamma\frac{k_{-3}}{k_2}u_1-\chi\sigma_2\right),
\label{selkove2}\\
&&\epsilon\frac{du_1}{d\theta}=u_2-\sigma_1u_1+\frac{k_{-3}}{k_2+k_{-1}}
(\sigma_2^\gamma(1-u_1-u_2)-u_1),\label{selkove3}\\
&&\epsilon\frac{du_2}{d\theta}=\sigma_1u_1-u_2.\label{selkove4}
\end{eqnarray}
Explicit expressions for the parameters $\epsilon$, $\nu$, $\eta$ and $\chi$
can be found in \cite{brechmann17} or \cite{keener09}. Formally setting 
$\epsilon=0$ in the equations (\ref{selkove3}) and (\ref{selkove4}) gives 
$u_2=\sigma_1 u_1$ and 
$u_1=\frac{\sigma_2^\gamma}{1+\sigma_2^\gamma+\sigma_1\sigma_2^\gamma}$ and
substituting these relations into the evolution equations for $\sigma_1$ and 
$\sigma_2$ gives
\begin{eqnarray}
&&\frac{d\sigma_1}{d\theta}=\nu-\left(
\frac{\sigma_1\sigma_2^\gamma}{1+\sigma_2^\gamma+\sigma_1\sigma_2^\gamma}\right),
\label{selkovlim1}\\
&&\frac{d\sigma_2}{d\theta}=\eta \left(
\frac{\sigma_1\sigma_2^\gamma}{1+\sigma_2^\gamma+\sigma_1\sigma_2^\gamma}
-\chi\sigma_2\right).\label{selkovlim2}
\end{eqnarray}
It can be shown that solutions of this system of two equations can be 
approximated by solutions of the four-dimensional system 
(\ref{selkove1})-(\ref{selkove4}) using geometric
singular perturbation theory \cite{kuehn15}. To do this we need to examine
the transverse eigenvalues. This means computing the matrix of partial 
derivatives of the right hand sides of the evolution equations for $u_1$ and 
$u_2$ with respect to the variables $u_1$ and $u_2$ and evaluating the
result for 
$\epsilon=0$. A computation shows (for details see \cite{brechmann17}) that
this matrix has determinant $\frac{k_{-3}}{k_2+k_{-1}}
(\sigma_1\sigma_2^\gamma+\sigma_2^\gamma+1)>0$ and trace 
$-\sigma_1-\frac{k_{-3}}{k_2+k_{-1}}(\sigma_2^\gamma+1)-1<0$. Hence both 
eigenvalues have negative real parts and the limit is well-behaved.

Selkov claims that starting from the assumptions of Higgins and supposing on
biological grounds that certain quantities are small leads to system 
(\ref{selkovlim1})-(\ref{selkovlim2}) with $\gamma=1$. This system has
the unfortunate property that it does not admit periodic solutions.
This can be proved using the fact that 
$\frac{1+\sigma_2^\gamma+\sigma_1\sigma_2^\gamma}{\sigma_1\sigma_2}$ is a
Dulac function.

Consider now the additional rescaling 
$x=\frac{\nu^{\gamma-1}}{\chi^\gamma}\sigma_1$, $y=\frac{\chi}{\nu}\sigma_2$
and $\tau=\left(\frac{\nu}{\chi}\right)^{\gamma}\theta$. In the equations 
for $\frac{dx}{d\tau}$ and $\frac{dy}{d\tau}$ with parameters $\nu$, $\chi$
and $\eta$ we can replace $\eta$ by 
$\alpha=\frac{\eta\chi^{\gamma+1}}{\nu^\gamma}$. Then the limit $\nu\to 0$ is 
regular. In this limit we obtain the system (\ref{selkov1})-(\ref{selkov2})
studied in this paper. This means that when the Selkov system has a periodic 
solution (which is then known to be hyperbolic and stable) the system 
(\ref{selkovlim1})-(\ref{selkovlim2}) will also have a periodic solution for
certain values of the parameters which is hyperbolic and stable. These are 
parameters for which $\chi$ is of order one, $\nu$ is small and $\eta$ is 
small. Geometric singular perturbation theory then allows us to conclude that 
the system (\ref{selkove1})-(\ref{selkove4}) possesses a periodic solution 
which is hyperbolic and stable for $\epsilon$ small.
 
Note that there is an alternative mathematical model of glycolysis due to 
Goldbeter and Lefever \cite{goldbeter72} which appears to explain some of 
the experimental observations better than that of Selkov. For a discussion
of this issue we refer to \cite{keener09}. The model of \cite{goldbeter72} 
exhibits intricate dynamical behaviour which has been investigated using
advanced methods of geometric singular perturbation theory in \cite{kosiuk11}. 
The Goldbeter-Lefever model was also studied mathematically in 
\cite{donofrio11}.

\vskip 10pt\noindent
{\it Acknowledgement} We thank Hussein Obeid for drawing our attention to the
possible relevance of quasihomogeneous directional blow-ups for this problem.

\end{document}